\documentstyle[epsfig,aps,prb,multicol]{revtex}
\begin{document}

\title{Rules for Minimal Atomic Multipole Expansion of Molecular
Fields}

\author{E.V. Tsiper$^{1,2,3*}$ and K. Burke$^1$}

\address{
$^1$Department of Chemistry \& Chemical Biology, Rutgers
University, Piscataway, NJ 08854\\
$^2$Center for Computational Material Science, Naval Research
Laboratory, Washington, DC 20375\\
$^3$School of Computational Sciences, George Mason University,
Fairfax, VA 22030}

\date{Novemebr 19, 2003}

\maketitle

\begin{multicols}{2}
 [ \begin{abstract}
 A non-empirical minimal atomic multipole expansion (MAME) defines
atomic charges or higher multipoles that reproduce electrostatic
potential outside molecules.  MAME eliminates problems associated with
redundancy and with statistical sampling, and produces atomic
multipoles in line with chemical intuition.
 \end{abstract} ]

The problem of representing the electrostatic potential outside a
molecule using atomic charges or higher atomic multipoles is very
important for understanding intermolecular forces.  Atomic partial
charges, an important part of chemical intuition, are defined in many
different ways for different purposes.  Chemically-derived (CD)
charges, such as Mulliken \cite{mulliken} or L\"owdin\cite{lowdin},
often describe molecular fields poorly.\cite{mulliken_fail} More
recent schemes partition molecular density into atomic regions, which
may or may not overlap.\cite{AIMs} Similar approaches have been
developed for solids.\cite{Boyer} Most attractive for our purposes are
potential-derived (PD) charges, which avoid representation of the
density by producing the `best' fit to the molecular potential
directly.\cite{momany,williams_1988} Atomic dipoles and
quadrupoles\cite{williams_1988} are often used to increase accuracy in
solvation problems\cite{martin_JPC} and force field
calculations.\cite{ren_2002} Induced atomic dipoles appear naturally
in electronic polarization of molecular solids\cite{pplus} to account
for the small part of molecular polarization that is due to the
deformation of atomic orbitals and is not captured by redistribution
of charges.

Computational schemes for PD multipoles such as Merz-Kollman
(MK),\cite{MK} CHelp\cite{CHelp}, or CHelpG\cite{CHelpG} differ mainly
in the sampling domain and the resulting atomic charges are strongly
method-dependent.\cite{sigfridsson} Worse still, PD methods often
yield atomic charges that are counter-intuitive, such as
negative charges on hydrogens in alkanes.\cite{williams_1994} Higher
multipoles only increase the redundancies inherent in
distributed multipole analysis, improving on the accuracy of the field
at the expense of instability in the multipole values.  The severity
of the problem can be somewhat reduced with SVD
techniques,\cite{svd,sigfridsson} or by introducing
restraints.\cite{RESP} Our approach does not use sampling and
eliminates redundancies before they appear.

We approximate the true molecular potential, $\phi({\bf r})$, as a sum
of multipoles of strength $q_k$ centered at nuclear positions ${\bf
r}_i$,

 \begin{equation}
 \phi({\bf r})\approx
 \phi_{\text{approx}}({\bf r})=\sum_i\sum_kq_k\phi_k({\bf r}-{\bf r}_i)
 \label{approx}
 \end{equation}
 where $\phi_k({\bf r})$ is the potential due to the $k$th multipole
of unit strength: $\phi_k({\bf r})=1/r$ for charge, $({\bf n}\cdot{\bf
r})/r^3$ for a dipole in the direction ${\bf n}$, and so on. Since
$\nabla^2\phi_{\text{approx}}=0$ everywhere except at ${\bf r}_i$, but
$\nabla^2\phi=4\pi\rho({\bf r})$, the atomic multipole expansion can
{\em only} be accurate in regions where $\rho({\bf r})\approx 0$.
Furthermore, $\phi$ on any closed surface $S$ on which $\rho=0$,
determines $\phi({\bf r})$ everywhere outside $S$.  We therefore
choose $S$ to be an isodensity surface, $\rho({\bf r})=f$, where $f$
is sufficiently small to ensure negligible charge beyond $S$, but with
sufficient potential on $S$ for a determining fit (Fig.~1).  We chose
$\phi_{\text{approx}}$ to minimize

 \begin{equation}
 \sigma^2=S^{-1}\oint_S dS\
   \left[\phi_{\text{approx}}({\bf r})-\phi({\bf r})\right]^2
 \label{sigma}
 \end{equation}
 over $S$ which leads to a system of linear equations
$\sum_kC_{mk}q_k=b_m$, where
 
\begin{mathletters}
 \begin{eqnarray}
 C_{mk}&=&S^{-1}\oint_S dS\ \phi_m({\bf r})\phi_k({\bf r}),\\
 b_m&=&S^{-1}\oint_S dS\ \phi_m({\bf r})\phi({\bf r}).
 \end{eqnarray}
 \label{integrals}
 \end{mathletters}
 Atomic multipoles defined in this way are fully rotationally
invariant, which is an issue with some PD schemes.\cite{CHelpG} The
error $\sigma$ can be compared to $\overline\phi$,

 \begin{equation}
 \overline\phi^2=S^{-1}\oint_S dS\ \phi^2({\bf r}).
 \label{phibar}
 \end{equation}

\centerline{\epsfig{file=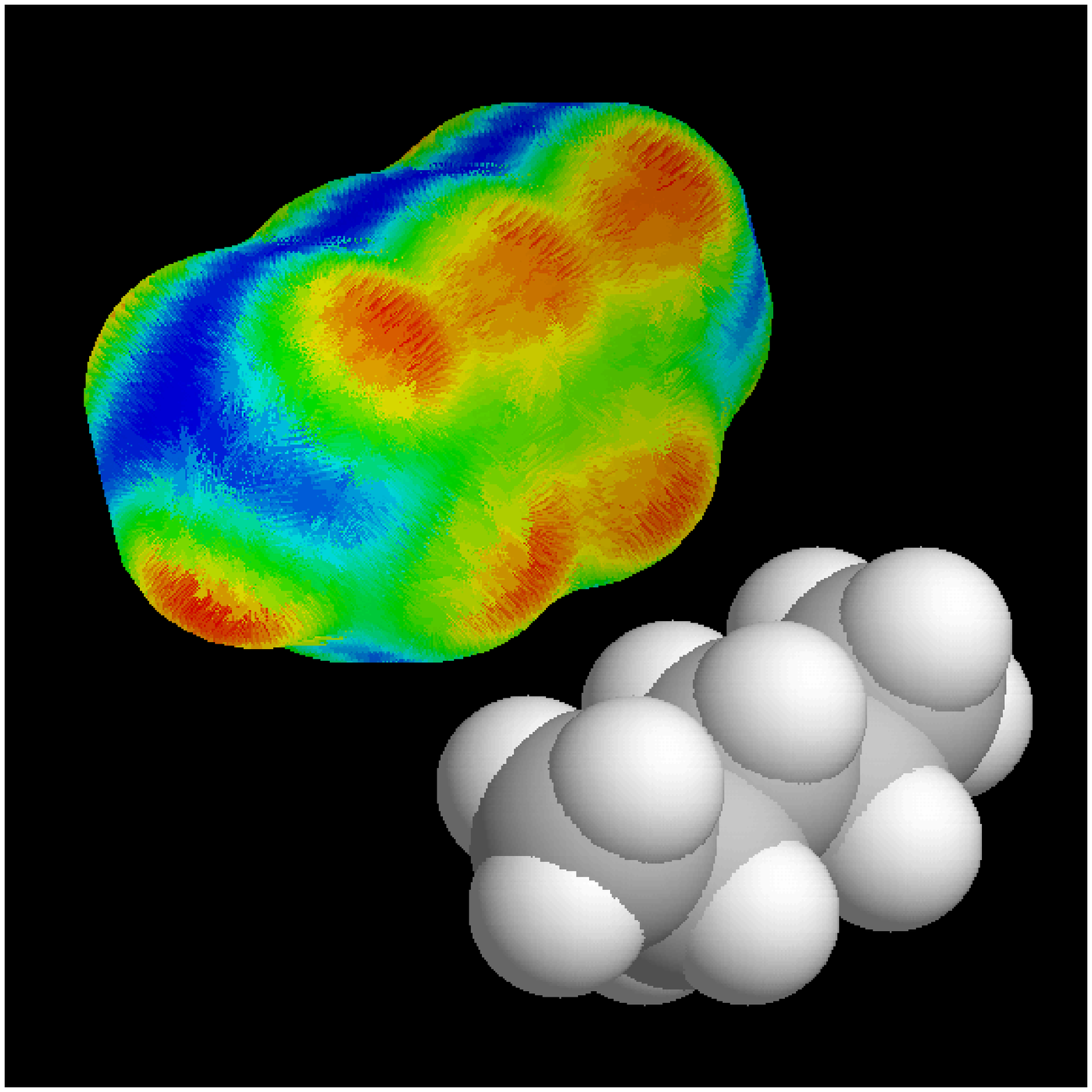,width=6cm}}

% \vskip 0.05 in
 {\small {\bf Fig.~1} Electrostatic potential over isodensity surface
$S$ of pentane.  Here $f=5\times10^{-4}$ au.~produces $S$ at $\sim1.4$
\AA\ from the hydrogens and leaves $-0.2e$ charge outside.  Missing
charge is negligible for $f=10^{-4}$, with $S$ at 1.8 \AA.}
\vskip 0.15 in

 The crucial issue remaining is the choice of a set of multipoles.  We
choose a minimal set, usually one scalar value per atom, and add
additional multipoles to describe lone pairs when necessary, based on
the Lewis structure.  This carefully chosen minimal atomic multipole
expansion (MAME) set avoids redundancies but is within $\sim1$ mH
everywhere beyond $S$.

We illustrate MAME with three molecules: n-pentane, which is a classic
example of difficulties encountered in PD schemes; glycine (standard
and zwitterion), as a typical application in biochemistry; and water,
to see how general MAME rules apply to a small polar molecule.  All
densities and potentials are produced on a cubic mesh by the Gaussian
98 program,\cite{gaussian} at the B3LYP/aug-cc-pVTZ level (6-311++G**
for pentane).  Surface integrals (\ref{integrals}) are computed by
triangulation of $S$.  The program runs within a few seconds, and is
available on request.

Figure 1 shows $\phi({\bf r})$ on $S$ for $n$-pentane.  Red spots
($\phi>0$) show an excess of positive charge near each hydrogen, but
all PD schemes tested yield some or all hydrogens negative.  Closer
inspection of Fig.~1 reveals that the positive regions occupy less
solid angle around hydrogens than would be produced by a positive
charge.  Such a potential is consistent with a {\em dipole} with a
negative charge pointing inwards.

Our first rule is therefore to assign a charge to all nuclei but
protons, to which we assign a dipole moment instead.  The hydrogen
atom is special as its sole electron participates in the bond, leaving
no electron density centered on the proton.  This unique property of
hydrogens is well-known in X-ray structure analysis, which
systematically underestimates the C---H bond lengths for this reason.

 \vskip 0.1 in
 {\small {\bf Table 1} Partial atomic charges in $n$-pentane.
$f=5\times10^{-4}$ au., error $\sigma$ as in (\ref{sigma}),
(\%)$=(\sigma/\overline\phi)$, $e\overline\phi=3.6$ mHartree
($1\text{mH}=27\text{meV}\sim kT$ at 300K). ``$\bbox{\mu}$'' indicates
atomic dipoles, ``$\mu_r$'' --- dipoles restricted along H---C bonds.}

 \vskip 0.05 in

\centerline{
\begin{tabular}{lc|c|cr}
\tableline
\tableline
 Method  & q(H), range & q(C), range & \multicolumn{2}{l}{$e\sigma$, mH (\%)} \\
\multicolumn{3}{l}{\ \ \ \ \ \ \ \ \ CD charges}\\
Mulliken             & $+0.11..+0.14$ & $-0.59..-0.11$ &\ 9.3 & (260)\\
ZINDO                & $+0.03..+0.04$ & $-0.15..-0.04$ &\ 3.5 & (99) \\
\multicolumn{3}{l}{\ \ \ \ \ \ \ \ \ PD charges}\\
CHelp                & $-0.04..+0.04$ & $-0.11..+0.15$ &\ 3.5 & (97) \\
CHelpG               & $-0.04..+0.04$ & $-0.16..+0.16$ &\ 3.2 & (87) \\
MK                   & $-0.03..+0.06$ & $-0.22..+0.13$ &\ 3.1 & (86) \\
\multicolumn{3}{l}{\ \ \ \ \ \ \ \ \ PD charges plus dipoles}\\
CHelp $+\bbox{\mu}$  & $-0.76..+0.10$ & $-0.66..+2.05$ &\ 2.8 & (78) \\
CHelpG $+\bbox{\mu}$ & $-0.32..-0.30$ & $+0.65..+0.86$ &\ 1.8 & (51) \\
MK $+\bbox{\mu}$     & $-0.27..-0.20$ & $+0.50..+0.64$ &\ 1.8 & (49) \\
\multicolumn{3}{l}{\ \ \ \ \ \ \ \ \ MAME}\\
charges              & $-0.01..+0.09$ & $-0.34..+0.13$ &\ 2.6 & (72) \\
$\bbox{\mu}$(H)      &$(\mu=0.07..0.09)$& $-0.01..+0.03$ &\ 0.5 & (15) \\
$\mu_r$(H)           &$(\mu=0.06..0.09)$& $-0.02..+0.01$ &\ 1.6 & (45) \\
\tableline
\tableline
\end{tabular}}

 \vskip 0.1 in

Mulliken charges are intuitively meaningful but produce large errors
in the potential (Table 1).  PD charges are negative on some hydrogens
and still give significant errors.  Adding dipoles reduces the
potential error, but at the cost of producing meaningless
multipoles.\cite{williams_1994} Our scheme with charges on all atoms
produces similar (though better) results, but we do far better (line
2) when the charges on hydrogens are replaced with dipoles.  All
dipoles come out similar in magnitude (numbers in brackets, in au.)
and point toward C within 20$^\circ$ of the H---C bond. The hydrogen
dipoles can be safely restricted to lie along the H---C bonds (last
line) with the accuracy still better than that of charges.  All
multipoles have reasonable values, including small charges on carbons.
Note that we have now described the field outside the molecule {\em
more accurately} than any existing scheme, with only {\em one}
parameter per nuclues (a charge on each carbon and a bond-directed
dipole on each hydrogen).

The same choice of multipoles yields a 1.05 mH error (=2\%) in the
glycine zwitterion, (NH$_3$)$^+$--CH$_2$--COO$^-$, down from 4\% with
charges alone and 4\%---6\% with standard PD schemes.  The glycine
zwitterion is highly polar with dipole $\mu=10.3$ D, which MAME
recovers within 0.1\% accuracy.

 \vskip 0.1 in
 {\small {\bf Table 2} MAME for glycine without and with lone pair
multipoles. $f=10^{-4}$ au., $e\overline\phi=15$ mH.}

 \vskip 0.05 in

\centerline{
\begin{tabular}{l|ccccc|c}
\tableline
\tableline
                 & NH$_2$ & CH$_2$ &  C$=$  &  $=$O  &  $-$OH & $e\sigma$, mH (\%) \\
\tableline 
$\mu_r$(H)       & $-.05$ & $-.03$ & $+.79$ & $-.54$ & $-.17$ & 4.1 (27) \\
\tableline 
$\mu_r$(H)       & $+.06$ & $+.11$ & $+.37$ & $-.67$ & $+.13$ & \\
\ \ $+\mu_r$(N,O)& $-.65$ &        &        & $+.19$ & $-.71$ & 1.6 (11) \\
\ \ $+\theta_r$(O)    &        &        &        & $-.53$ & $-.87$ & \\
\tableline
\tableline
\end{tabular}}

 \vskip 0.1 in

Table 2 lists MAME results for glycine in its standard form,
NH$_2$--CH$_2$--COOH, and illustrates the need for special treatment
of lone pairs.  In the zwitterion, the NH$_3$ group is well-described
by a charge on N and three dipoles on hydrogens, similar to methyls in
pentane.  The NH$_2$ group in glycine lacks one site, but has extra
electron density associated with the lone pair.  We thus assign a
dipole moment to N, in addition to its charge, restricted along the
$sp^3$ direction of the lone pair.

Similarly, each oxygen has two lone pairs.  Two dipoles for the two
lone pairs sum to just one dipole along the symmetry axis, leading to
only one variational parameter.  The potential of this single dipole,
however, is axially symmetric, whereas the potential around the oxygen
deviates from axial symmetry due to the particular orientation of the
lone pairs.  Such a deviation can be accounted for with a quadrupole
moment on the oxygen.  The finite system of charges sketched in the
inset in Fig.~2 shows what is needed.  Computing a multipole expansion
of three charges we describe two lone pairs with two scalar
parameters, a dipole $\mu_r$ restricted along the symmetry axis, and a
{\em restricted quadrupole} $\theta_r$ which has angle $\beta$ as a
fixed parameter ($\beta=120^\circ$ for O$=$ and $109.7^\circ$ for O--,
due to $sp^2$ and $sp^3$ hybridization respectively).  $\theta_r$ and
$\mu_r$ are chosen negative with ``$-$'' pointing outside.  Table 2
shows a clear advantage of such a multipole set.

To make the definition more transparent, $\theta_r$ can be expressed
in the conventional form as a carefully crafted combination of
$\theta_{zz}=\theta_r(3\cos^2\beta/2-1)$,
$\theta_{yy}=\theta_r(3\sin^2\beta/2-1)$, $\theta_{xx}=-\theta_r$, and
$\theta_{xy}=\theta_{xz}=\theta_{yz}=0$, which depend on a single
scalar parameter $\theta_r$.  Here $z$ is along the symmetry axis, and
the lone pairs are in the $yz$-plane.  A single restricted quadrupole
of strength $\theta_r$ creates the potential

\begin{equation}
\phi_{quad}({\bf r})=\theta_r
\frac{3({\bf r}\cdot{\bf n}_1)^2+3({\bf r}\cdot{\bf n}_2)^2-2r^2}
  {2r^5},
\label{phi_d}
\end{equation}
 where ${\bf n}_1$ and ${\bf n}_2$ are the directions of the lone
pairs.

For the zwitterion, MAME does not require $\theta_r$ on the oxygens,
because of the resonance.  Lone pairs in the $sp^2$ and $sp^3$
configurations lie in perpendicular planes, virtually destroying any
asymmetry.

MAME accuracy improves away from the molecule (Fig.~2).  If $f$ is too
large ($10^{-3}$), there is a net charge inside $S$ which strongly
affects the asymptotic behavior.  This can be repaired by fixing the
total charge using a Lagrange multiplier.\cite{svd} Figure 2
demonstrates MAME's insensitivity to choice of $f$, provided the total
charge is correct.

\centerline{\epsfig{file=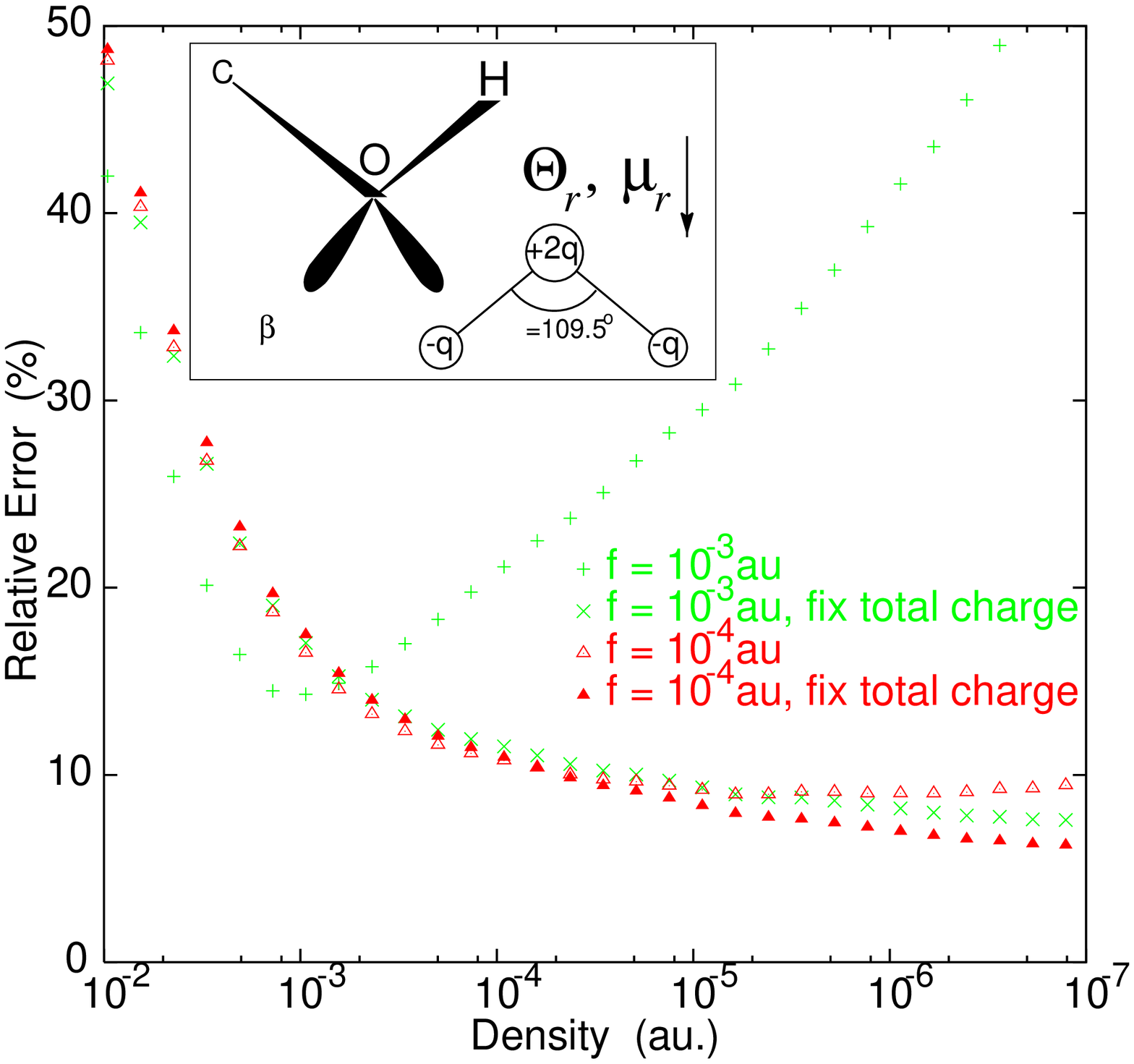,width=7cm,angle=-90}}

 \vskip 0.05 in
 {\small {\bf Fig.~2} MAME accuracy away from glycine. Ratio of
$(\phi_{\text{approx}}-\phi)$ to $\phi$, both square-averaged over
points with a given density $\rho$, which is an inverse measure of
distance.  The inset illustrates two lone pairs on oxygen represented
with a combination of restricted dipole $\mu_r$ and quadrupole $\theta_r$.}
 \vskip 0.1 in

Last, we analyze the water molecule.  The first rule leads to two
dipoles on hydrogens pointing along the bonds, sketched in Fig.~3.
The charge on oxygen is zero because the molecule is neutral.  The
dipoles are equal due to symmetry and require no calculation since
their vector sum must yield the dipole moment of water, 1.847D
(B3LYP/aug-cc-pVTZ value).  This already reduces the error to 21\%,
from 45\% with a single dipole on oxygen.  The two distributed dipoles
yield $\Theta_{xx}-\Theta_{yy}=4.06$ D\AA\ for the quadrupole moment
of water, whereas experiment \cite{verhoeven} gives 5.12 D\AA.  We
note that $\Theta_{xx}-\Theta_{yy}$ is the only invariant combination
of quadrupole components, since the finite dipole makes them dependent
on the center of coordinates.

We next add $\mu_r$ and $\theta_r$ multipoles on the oxygen to
describe the correction due to lone pairs, and we find excellent
accuracy $e\sigma=0.59$ mH ($<3$\%) on and beyond $S$ (1.56\AA\ from H
and 2.11\AA\ from O, $f=10^{-4}$).  The $sp^3$ choice of
$\beta=109.5^\circ$ in $\theta_r$ is crucial: accuracy deteriorates
dramatically (to 12\%) when $\beta$ is changed to e.g. $180^\circ$
($\theta_r$ replaced with $\theta_{yy}$).  The $sp^3$ description of
the oxygen lone pairs is appropriate due to invariance under unitary
rotations of occupied orbitals.

\centerline{\epsfig{file=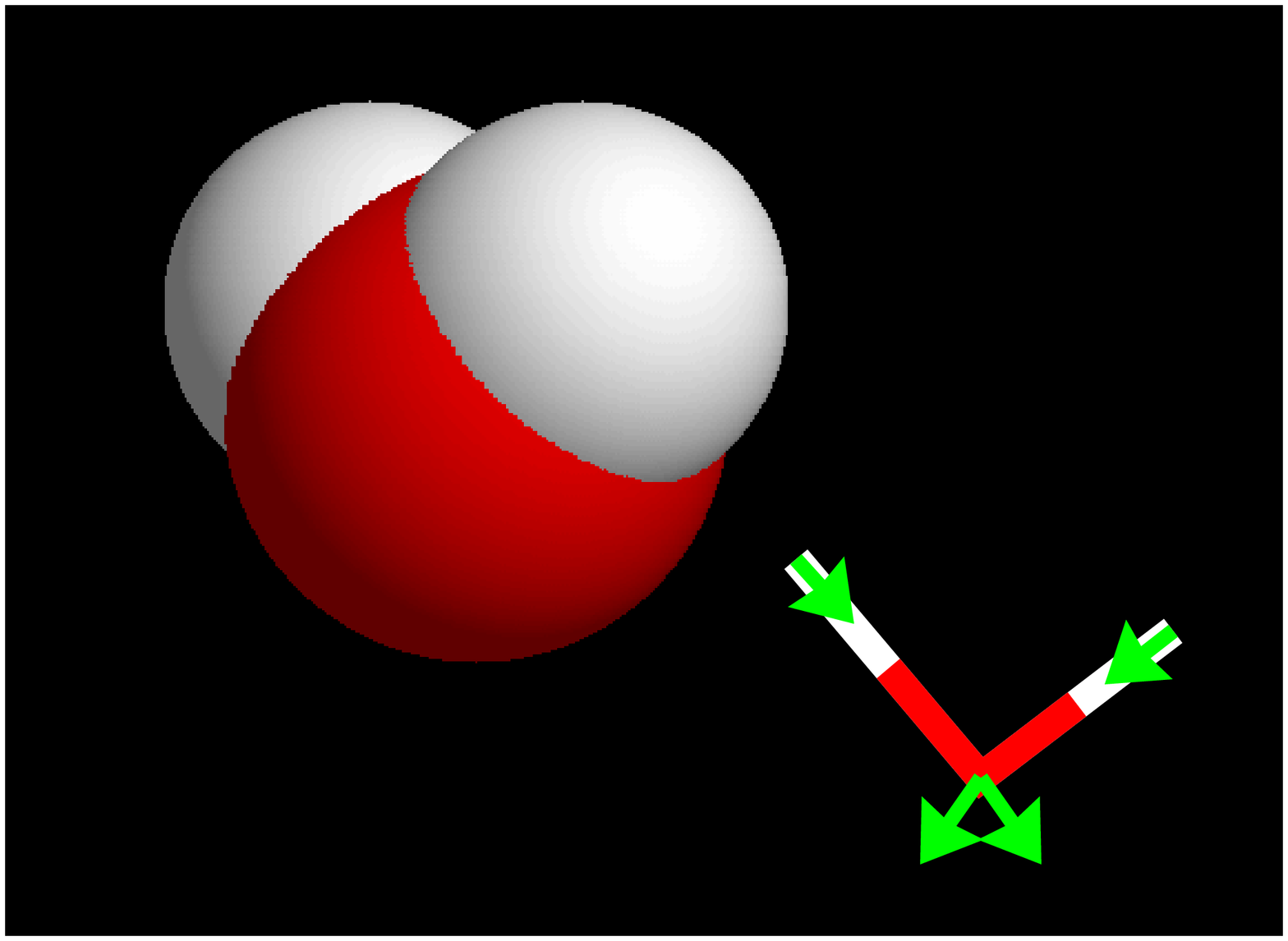,width=6cm,angle=-90}}

 \vskip 0.05 in
 {\small {\bf Fig.~3} Schematic MAME representation of water.}
 \vskip 0.1 in

In conclusion, molecular fields are represented to chemical accuracy
with a minimal set of atomic multipoles carefully chosen based on the
Lewis structure of the molecule.  All H atoms are represented as
dipoles.  Lone pairs are treated with extra multipoles, avoiding
additional off-nuclear expansion sites.\cite{dixon} The scheme yields
multipole values that conform to chemical intuition, are unique, fully
rotationally-invariant and free of sampling errors.

We thank K. Krogh-Jespersen, Z.G. Soos, M. Lobanov, and R. Gaudoin for
enlightening discussions.  This work was funded by AFOSR and the New
Jersey Commision on Science and Technology.

\end{multicols}
\end{document}